\documentclass[11pt, two column, journal]{IEEEtran}

\ifCLASSINFOpdf

\else

\fi

\usepackage{amsfonts}

\usepackage{amssymb}

\usepackage{amsmath}

\usepackage{amscd}

\usepackage{amsthm}

\usepackage{fancyvrb}

\usepackage{epsfig}

\usepackage{mathtools}

\usepackage{epsfig}

\usepackage{bbm}

\usepackage{fullpage}

\usepackage{mathrsfs}

\usepackage{alltt, url}

\usepackage{graphicx}

\usepackage[boxed,algosection,linesnumbered]{algorithm2e}

\begin{document}

%



\title{A Near Optimal Approach for Symmetric Traveling Salesman Problem in Euclidean Space}

\author{\IEEEauthorblockN{Wenhong Tian$^{1,2}$, Chaojie Huang$^{2}$, Xinyang Wang$^{2}$\\}

\IEEEauthorblockA{Chongqing Institute of Green and Intelligent Technology, Chinese Academy of Sceinces ,\\ School of Information and Software Engineering\\
University of Electronic Science and Technology of China, China \\
Email: tianwenhong@cigit.ac.cn}}

\maketitle

\begin{abstract}

The traveling salesman problem (TSP) is one of the most challenging NP-hard problems. It has widely applications in various disciplines such as physics, biology, computer science and so forth. The best known approximation algorithm for Symmetric TSP (STSP) whose cost matrix satisfies the triangle inequality (called $\triangle$STSP) is Christofides algorithm which was proposed in 1976 and is a $\frac{3}{2}$-approximation. Since then no proved improvement is made and improving upon this bound is a fundamental open question in combinatorial optimization.
In this paper, for the first time, we propose Truncated Generalized Beta distribution (TGB) for the probability distribution of optimal tour lengths in a TSP. We then introduce an iterative TGB approach to obtain quality-proved near optimal approximation, i.e., (1+$\frac{1}{2}(\frac{\alpha+1}{\alpha+2})^{K-1}$)-approximation where $K$ is the number of iterations in TGB and $\alpha (>>1)$ is the shape parameters of TGB. The result can approach the true optimum as $K$ increases. \\
\end{abstract}
%

\begin{IEEEkeywords}

Symmetric  traveling salesman problem (STSP); Triangle inequality; Random TSP in a unit square; TSPLIB instances;Approximation ratio; k-opt; Computational complexity

\end{IEEEkeywords}






%



\IEEEpeerreviewmaketitle

\section{Introduction}

\indent
The TSP is one of most researched problems in combination optimization because of its importance in both academic need and real world applications. For surveys of the TSP and its applications, the reader is referred to [Cook,2012][An et al., 2012][Vygen, 2012] and references therein.

After 39 years, Christofides' $\frac{3}{2}$-approximation algorithm [Christofide, 1976] still keeps the best performance guarantee known for the symmetric traveling salesman problem satisfying triangle inequality ($\triangle$STSP), and improving upon this bound is a fundamental open question in combinatorial optimization, see [Cook, 2012][Gutin et al., 2002] and references therein. [Vygen, 2012] also provides a detailed survey on new approximation algorithms for the TSP. [Johnson et al., 1998] provide a complete comparative study on the local optimization methods for TSP. [Cook, 2012] introduces the TSP from history to the state-of-the art. David Johnson [2014] discusses the importance and applications of random TSP. 

[Gharan et al., 2011] give a ($\frac{3}{2}$-$\epsilon$)-approximation for a tiny $\epsilon>$0 using a probabilistic analysis, as [Klarreich, 2013] reports that the new progress is a 49.99...96$\%$ (totally 46 nines replaced by ``..." ) over the optimum, a tiny margin for ``graphical" traveling salesman problems. Notice that a very small percentage improvement may also be of great impact to the total length of large TSP instances.

[Reiter and Rice 1966] study the cost distribution of local optima under a gradient maximizing search in 39 integer programming problems. Their results suggest that the local optima follow a Beta distribution.
[Golden 1978] examines six problems from the TSPLIB archive [15]. The resulting estimates of optimal solutions are compared to the best solution found by the Lin-Kernighan algorithm [13]. The authors found that the Beta distribution is ``a more appropriate distribution'' than the Weibull distribution.

Recently, [Vig and Palekar 2008], apply sampling techniques similar to Golden, and use the Lin-Kernighan algorithm to find optimal tour costs. The authors estimate raw moments from the one to four of the probability distribution of optimal tour lengths. They use these estimates to fit various candidate distributions including the Beta, Weibull and Normal cases. Vig and Palekar conclude that the Beta distribution yields the best fit.

More recently, [Stuffle 2009] provide exact solutions to compute the mean, variance, the third and fourth central moment of all tour lengths. The computational complexity of computing variance, the third and fourth central moment is respectively $O(n^2)$,$O(n^4)$ and $(n^6)$ where $n$ is the number of nodes in a TSP. 

A typical probability distribution of all tour lengths for a random TSP in a square unit is shown in Fig. 1 where the total node number is 12. An example of TSPLIB Burma14 is shown in Fig.2. Similar results are observed for different total number of cities for which all tour lengths can be obtained.\\
The organization of remaining parts of this paper is: our major contributions are summarized in Section II, our methods are introduced in Section III, and Conclusions and future work are discussed in Section IV.

\begin{figure} [htp!]

\begin{center}


{\includegraphics [width=0.4\textwidth,angle=-0] {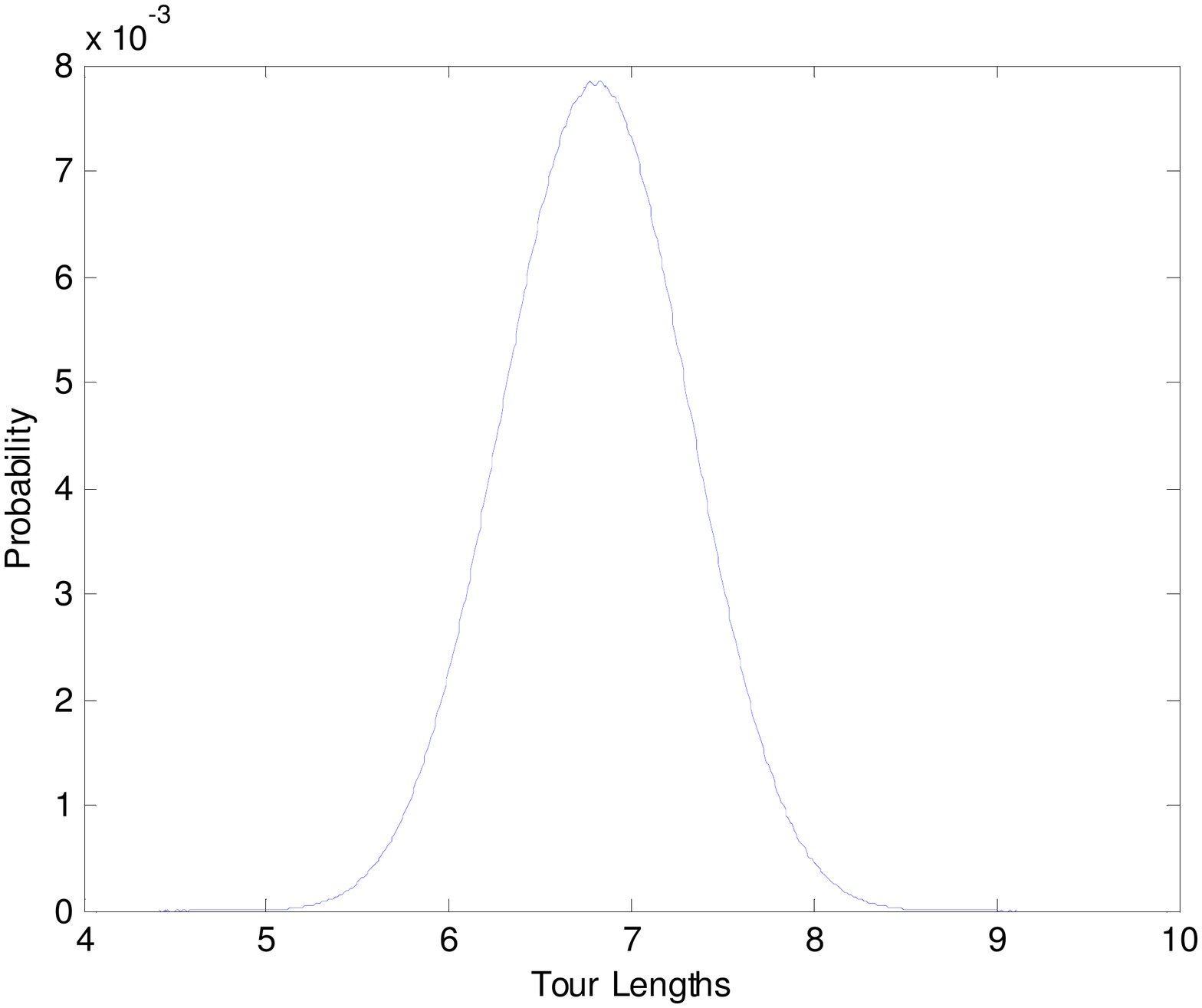}}


\caption{The probability density of random 12-node TSP in a squre unit }

\end{center}

\end{figure}

\begin{figure} [htp!]

\begin{center}


{\includegraphics [width=0.4\textwidth,angle=-0] {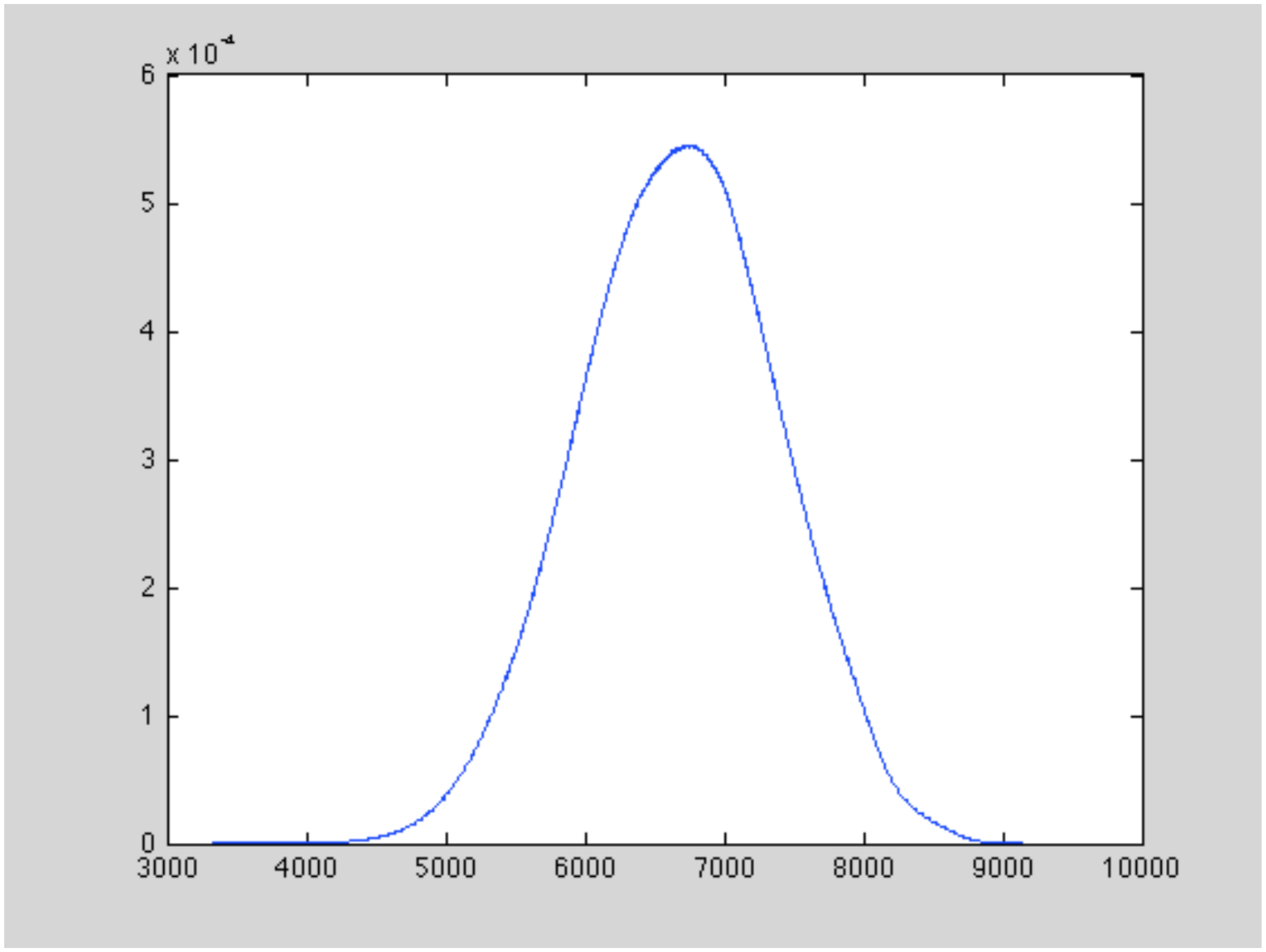}}


\caption{The probability density of all tour lengths in Burma14.tsp with 14 nodes }

\end{center}

\end{figure}

\section {Results}
The main contributions of this work can be summarized as follows:


\begin{itemize}

\item We propose Generalized Beta (GB) distribution as the probability density function of all tour lengths distribution in a symmetric TSP in Euclidean space (ESTSP), the four parameters of the GB can be computed from given ESTSP data directly.

\item For the first time, we introduce an iterative Truncated GB (TGB) closed-form solution to obtain (1+$\frac{1}{2}(\frac{\alpha+1}{\alpha+2})^{K-1}$)-approximation for a STSP where $K$ is the number of total iterations in TGB, and $\alpha (>>1)$ is the shape parameter of TGB and can be determined once the TSP instance is given. The result can approach the true optimum as $K$ increases.


\end{itemize}


\section{Methods}

Firstly, a problem formulation and some preliminaries are provided in this section.

\subsection{Problem Formulation}


Consider the $n$-node TSP defined in Euclidean space. This can be represented on a complete graph $G$= ($V,E$) where $V$ is the set of vertices and $E$ is the set of edges. The cost of an edge ($u$, $v$) is the Euclidean distance ($c_{uv}$) between $u$ and $v$. Let the edge cost matrix be $C[c_{ij}]$ which satisfies the triangle inequality.

\textbf{Definition 1.} Symmetric TSP (STSP) is TSP in Euclidean distance (called ESTSP) and the edge cost matrix $C$ is symmetric.

\textbf{Definition 2.} $\triangle$STSP is a STSP whose edge costs are non-negative and satisfies the triangle inequality, i.e., for any three distinct nodes (not necessary neighboring) ($i, j, k$), $(c_{ij}$+$c_{jk}) \geq c_{ik}$.

%
%
%

\textbf{Definition 3.} TSP tour. Given a symmetric graph $G$ in 2-dimensional Euclidean distance and its distance matrix $C$ where $c_{ij}$ denote the distance between node $i$ and $j$ (symmetrically). A tour $T$ has length

\begin{equation}
L=\sum_{k=0}^{N-1} c_{T(k),T(k+1)}
\end{equation}

where $N$ is the total number of nodes in $G$ and $T(N)$=$T(0)$ so that a feasible tour is formed.\\
\textbf{Definition 4.} The approximation ratio of an algorithm. The ratio is the result obtained by the algorithm over the optimum (abbreviated as OPT in this paper).

\textbf{Observation 1.} The probability density function of all tour lengths in an ESTSP can be modelled by a Generalized Beta (GB) distribution.

This is observed in Fig. 1 and other ESTSPs for which we can obtain all tour lengths. More results are provided in next section. This is also validated and shown in [Vig 2008], where a scaled Beta distribution is applied with scaled mean and scaled variance. The author validated estimated results by Anderson-Darling (A-D) test and Kolmogorov-Smirnov (K-S) test for random TSP. They use these estimates to fit various candidate distributions including the Beta, Weibull and Normal cases, and conclude that the (scaled) Beta distribution yields the best fit.

We further propose a Generalized Beta (GB) distribution.
The probability density function (pdf) of GB is defined as
\begin{equation}
f(x,\alpha,\beta, A, B)=\frac{(x-A)^{\alpha-1}(B-x)^{\beta-1}}{Beta(\alpha,\beta)}\label{eq:GBDensity1}
\end{equation}
where $Beta(\alpha,\beta)$ is the beta function
\begin{equation}
Beta(\alpha,\beta)=\int_{0}^{1}t^{\alpha-1}(1-t)^{\beta-1}dt \label{eq:beta1},
\end{equation}
$A$ and $B$ is the lower bound and upper bound respectively, $\alpha>0$, $\beta>0$, see [Hahn and Shpiro, page 91-98,126-128,1967]. For TSP, $A$ and $B$ represents the minimum and maximum tour length respectively.

The four central moments, mean ($\mu$), variance, skewness and kurtosis of the Generalized Beta distribution with parameters ($\alpha$,~$\beta$, $A$, $B$) are given by:

\begin{equation}
\mu=A+(B-A)\frac{\alpha}{\alpha+\beta}\label{eq:BetaMean}
\end{equation}
\begin{equation}
Var=(B-A)^2\frac{\alpha\beta}{(\alpha+\beta)^{2}(\alpha+\beta+1)}\label{eq:BetaVariance1}
\end{equation}

\begin{equation}
Skewness=\frac{2(\beta-\alpha)\sqrt{1+\alpha+\beta}}{\sqrt{\alpha+\beta}(2+\alpha+\beta)}\label{eq:BetaSkewness}
\end{equation}
and Kurtosis
\begin{equation}
\frac{6[\alpha^{3}+\alpha^{2}(1-2\beta)+\beta^{2}(1+\beta)-2\alpha\beta(2+\beta)]}{\alpha\beta(\alpha+\beta+2)(\alpha+\beta+3)}
\end{equation}

The standard deviation is then given by
\begin{equation}
\sigma=\sqrt{Var} \label{eq:StandDeviation}
\end{equation}
Once four central moments are known, or any four parameters of ($A$, $B$,~$\mu$, Var, skewness, kurtosis) are given, then four parameters of GB, i.e., ($\alpha$,~$\beta$,~$A$,~$B$) can be determined easily from the four moments match using Eqns.(4)-(7).
When the problem size is not large, the four central moments can be computed exactly using methods proposed in [Stuffle 2009]. As the problem size increases, we can find any four parameters of ($A$,~$B$,~$\mu$,~Var,~skewness) firstly, then find four parameters of GB, i.e., ($\alpha$,~$\beta$,~$A$,~$B$).

For medium or large size problem, currently it is not easy to find the fourth central moment. However, the lower bound (A) can be easily computed by LKH code [14]. So in the following sections, we find four values ($A$, mean ($\mu$),~variance, skewness) firstly, and then compute other parameters ($B$, $\alpha$, $\beta$). Firstly we introduce a method to compute maxTSP ($B$) [Gutin et al.,2002].\\
\textbf{Definition 5.} maxTSP. The maximum tour length (B) is obtained using LKH where each edge cost ($c_{ij})$ is replaced by a very large value ($M$) minus the original edge cost, i.e., ($M$-$c_{ij}$). $M$ can be set as the maximum edge cost plus 1.\\
Since the characteristics of random TSP and TSPLIB instances are different in Euclidean space, we introduce the GB as the probability density function for them separately.
\subsection {GB as the Probability Density Function for Random TSP}


For medium size random TSP problems with $n$ varying from 20 to 100, we can obtain four central moments easily [Suffle 2009] and apply four moments match to find four parameters for GB. After obtaining four parameters, we then use linear regression to find closed-form solution to $(\alpha,~\beta)$ of GB for random TSP. For $n$=20 to $n$=100, and we find that
\begin{equation}
\alpha(n)=1.9197n-32.166, R^2=0.9994 \label{eq:alpha1}
\end{equation}
\begin{equation}
\beta(n)= 1.1168n-15.854, R^2=0.9982 \label{eq:beta1}
\end{equation}
\begin{equation}
A(n)= 0.6932\sqrt{n}+0.8029, R^2=0.9956 \label{eq:mean}
\end{equation}
\begin{equation}
B(n)= 0.7649n-0.6393, R^2=0.998 \label{eq:mean}
\end{equation}
where $R^2$ is a measure of goodness-of-fit with value between 0 and 1, the larger the better. We observe that Eqns.(9)-(12) are highly accurate by extensive computation results.


\textbf{Observation 2.} The relative errors between estimated results (maxTSPs) by GB and LKH results are within $6.5\%$ for random TSP.
\\

The relative error is defines as (EstimatedValue-OPT)/OPT$\times 100\%$. We conduct tests for $n$=100 to $n$=500. The results are shown in Fig.3 where LKH is used to obtain maxTSP (B) results. 
We can observe that the relative error between our results and LKH are within 6.5$\%$ off the true optimums, with an average below $5\%$. Table 1 shows four parameters of GB for random TSPs with $n$ varying from 90 to 99.

\begin{figure} [htp!]

\begin{center}


{\includegraphics [width=0.4\textwidth,angle=-0] {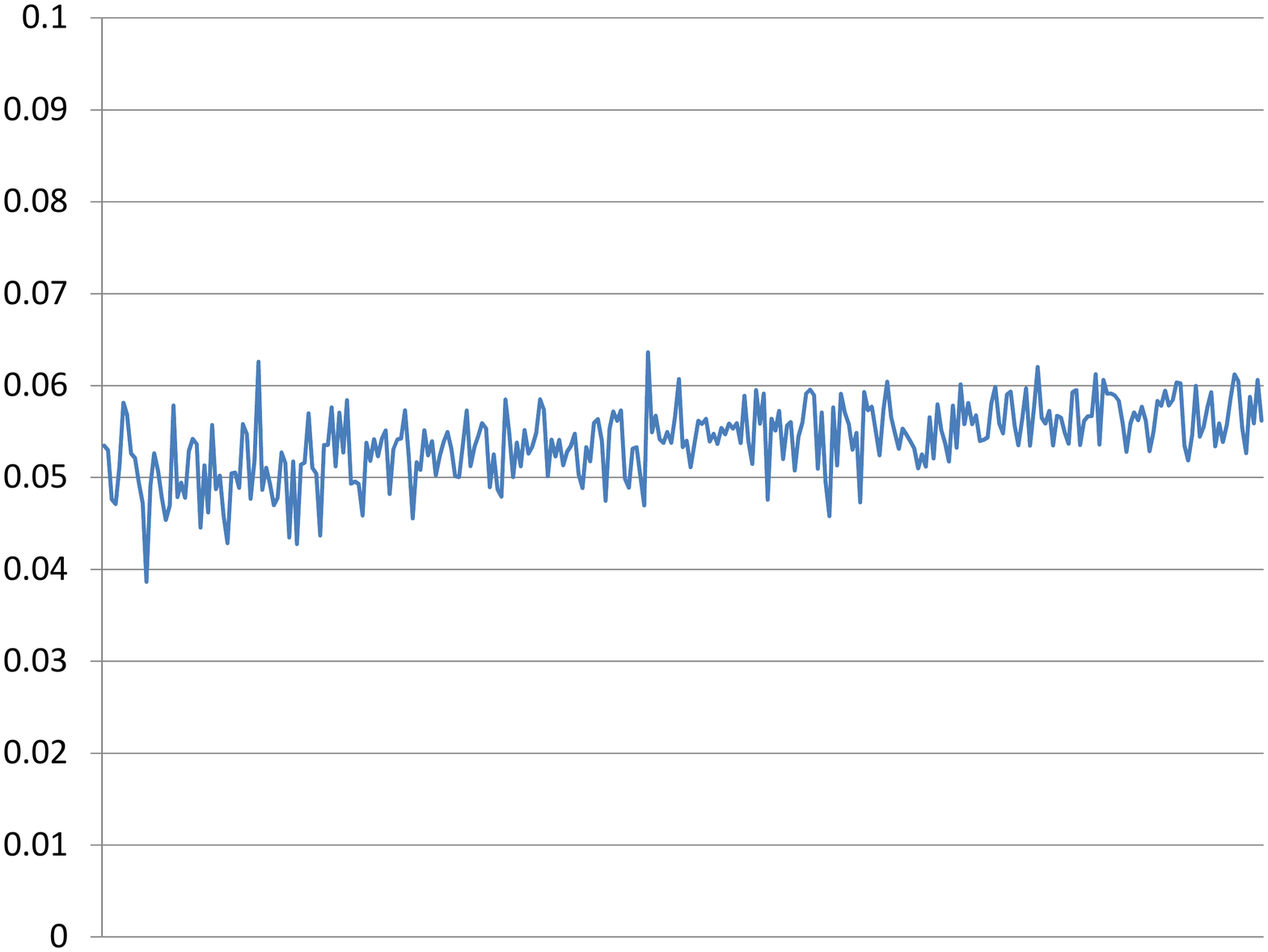}}


\caption{The relative error between estimated results by GB and OPT by LKH for Random TSP $n$=200 to 500}

\end{center}

\end{figure}

\subsection {GB as the Probability Density Function for TSPLIB Instances in Euclidean Distance}

Firstly, we show an example of Burma14.tsp, which we can permute its all tours and find that minimum tour length (A) is 3233 and maximum tour length (B) is 9139, Mean ($\mu$)=6679, Variance=503064, Skewness=-0.0632, Kurtosis=2.7972. Fig. 4 shows exact result (in black color) by permuting all tour lengths and estimated probability density function (in green color) by four central moment match with ($A$=3233, $B$=9579, $\alpha$=13.96,$\beta$=11.79). It can be observed that two results match very well.
\begin{figure} [htp!]

\begin{center}


{\includegraphics [width=0.4\textwidth,angle=-0] {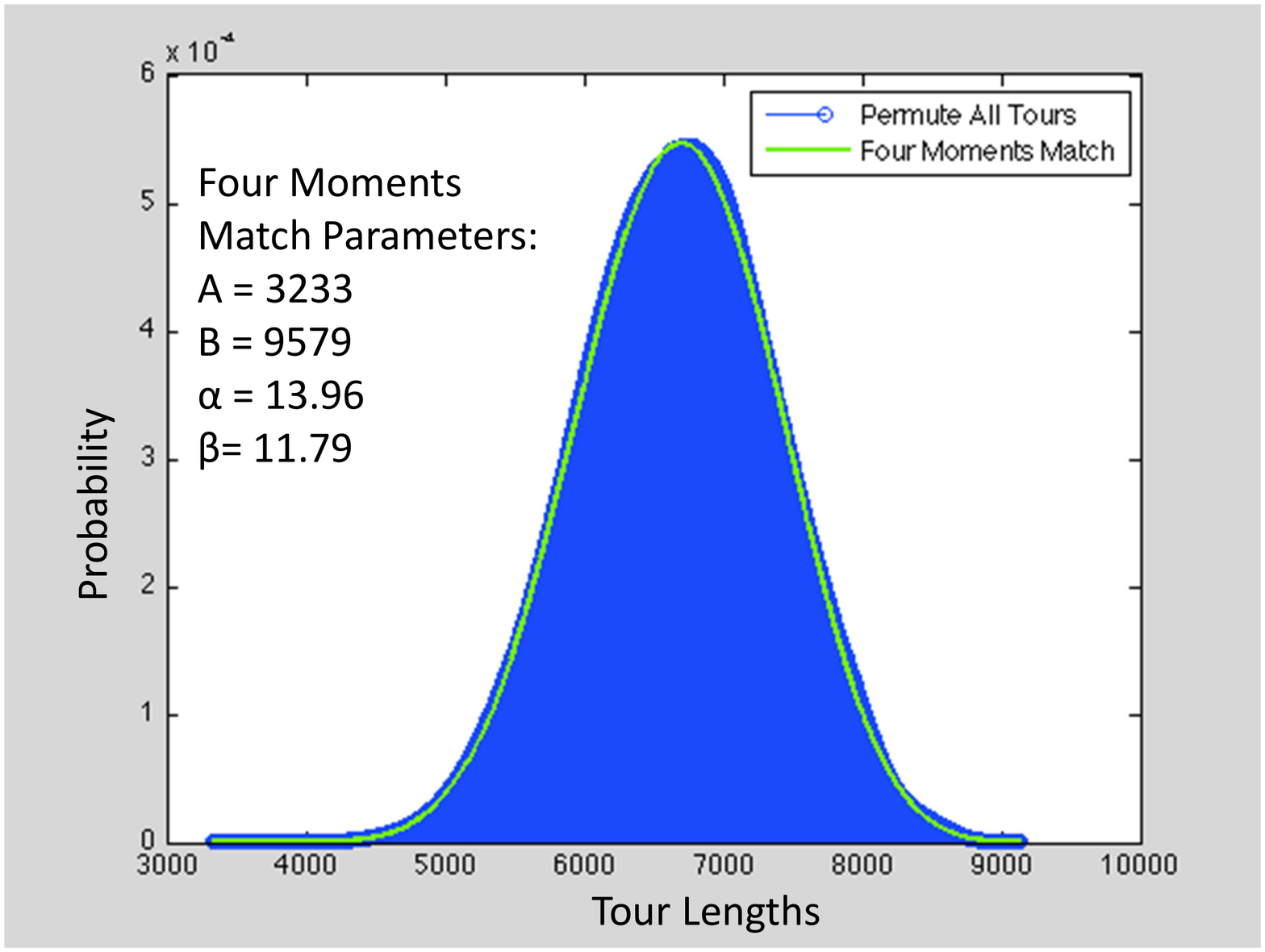}}


\caption{The exact (Permute All Tours) and estimated (Four Moments Match) probability distribution of Burma14.tsp }

\end{center}

\end{figure}
\\
\textbf{Observation 3.} The relative error between the estimated maximum tour lengths by GB and LKH results is below 7$\%$ for medium size TSPLIB instances, with an average below $5\%$.

We conduct tests by set $n$=14 to $n$=52 for which the four central moments can be easily computed and are given in [Stuffle 2009].
Fig. 5 shows the relative error between estimated results by GB and OPT by LKH.

\begin{figure} [htp!]

\begin{center}


{\includegraphics [width=0.4\textwidth,angle=-0] {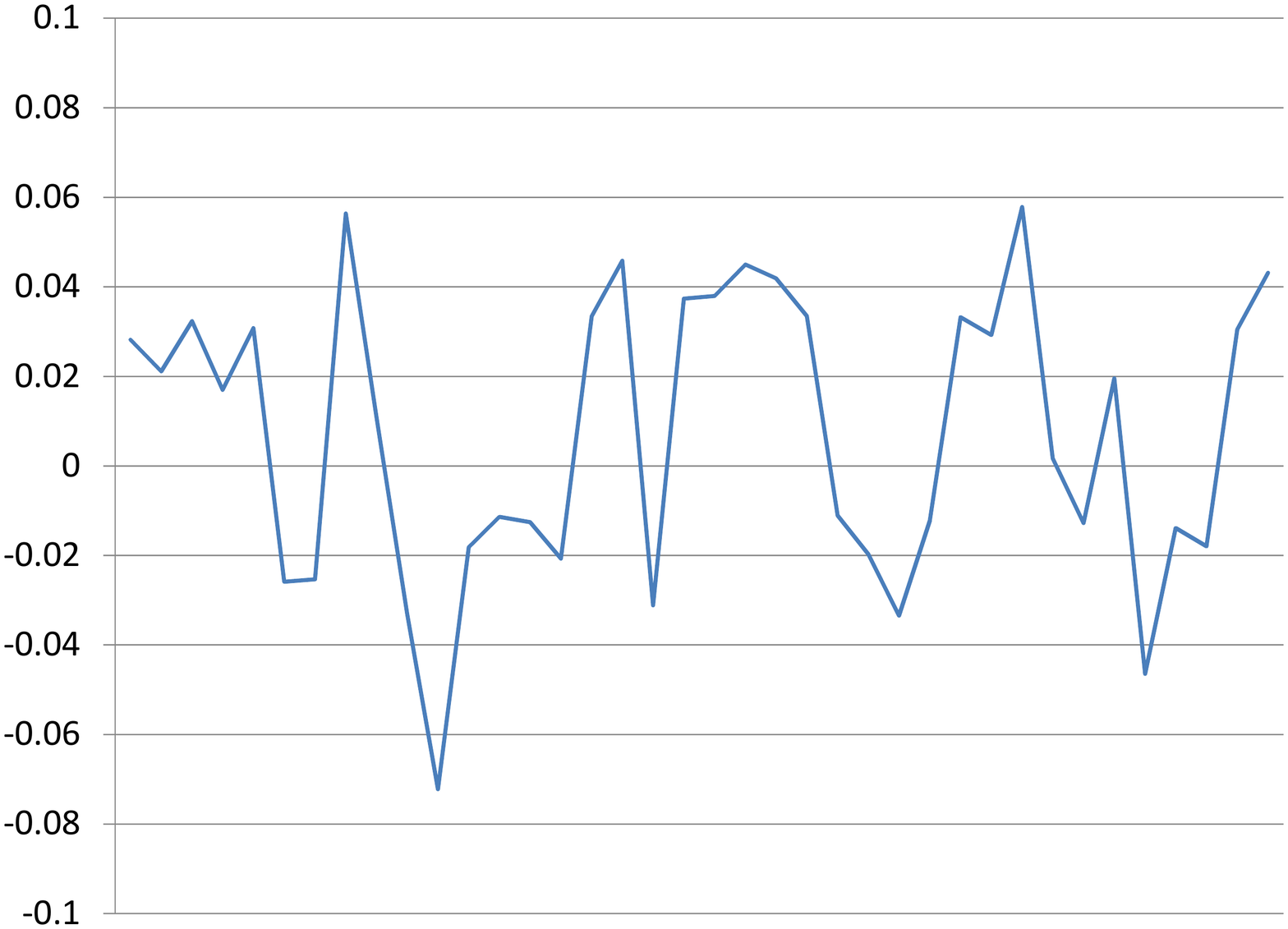}}


\caption{The relative error between estimated results by GB and OPT by LKH for TSPLIB instances for $n$ from 14 to 52}

\end{center}

\end{figure}

%
%
%
%
%
%
%

\begin{table}[ht]
\caption{Four parameters of GB for some random TSP instances } 
\centering 
\begin{tabular}{c c c c c} 
\hline\hline 
n & A(OPT) & B & $\alpha$ & $\beta$ \\ [0.5ex] 
\hline 
90& 7.18& 65.75& 179.68& 96.49\\
91& 7.15& 69.49& 180.38& 96.30\\
92& 7.01& 69.18& 185.51& 108.55\\
93& 8.00& 71.24& 182.22& 100.63\\
94& 7.73& 68.58& 175.38& 94.00\\
95& 7.77& 69.53& 181.09& 105.93\\
96& 7.40& 73.10& 204.60& 115.61\\
97& 8.00& 74.01& 186.85& 107.33\\
98& 7.73& 76.44& 208.67& 123.96\\
99& 7.54& 74.95& 203.46& 109.03\\[1ex]
\hline 
\end{tabular}
\label{table:nonlin} 
\end{table}

In Table 2 we show four parameters of GB for TSPLIB with $n$ varying from 14 to 52.
\begin{table}[ht]
\caption{Four parameters of GB for some TSPLIB instances } 
\centering 
\begin{tabular}{c c c c c} 
\hline\hline 
TSPLIB & A(OPT) & B & $\alpha$ & $\beta$ \\ [0.5ex] 
\hline 
burma14& 3323& 9139& 13.97& 11.79\\
ulysses16& 73.98& 180.52& 10.24& 6.52\\
gr17& 2085& 6160& 19.22& 10.60\\
gr21& 2707& 10680& 32.95& 19.80\\
ulysses22& 75.3& 241.50& 17.52& 12.79\\
gr24& 1272& 4929& 51.55& 27.21\\
fri26& 937& 3681& 28.87& 16.91\\
bayg29& 1610& 6654& 42.17& 26.42\\
bays29& 2020& 8442& 45.14& 27.52\\

\hline 
\end{tabular}

\label{table:nonlin} 
\end{table}


\subsection{Truncated Generalized Beta Distribution Based on Christofides Algorithm}

Next, we introduce our algorithm, \textbf{Truncated Generalized Beta distribution Based on Christofides Algorithm (TGB)}. TGB algorithm performs in seven steps: \\

\begin{itemize}

\item

(1). Finding the minimum spanning tree $MST$ of the input graph $G$ representation of metric TSP;

\item

(2). Taking $G$ restricted to vertices of odd degrees in $MST$ as the subgraph $G^{*}$; This graph has an even number of nodes and is complete;

\item

(3). Finding a minimum weight matching $M^{*}$ on $G^{*}$;

\item

(4). Uniting the edges of $M^{*}$ with those of the $MST$ to create a graph $H$ with all vertices having even degrees;

\item

(5). Creating a Eulerian tour on $H$ and reduce it to a feasible solution using the triangle inequality, a short cut is a contraction of two edges ($i, j$) and ($j, k$) to a single edge ($i, k$); 
\item
(6). Applying Christofides algorithm  to a ESTSP forms a truncated GB (TGB) for the probability density function of optimal tour lengths, with expectation (average) value at most 1.5OPT-$\epsilon$, where $\epsilon$ is a very small value; Applying $k$-opt to the result of Christofides algorithm forms another TGB for probability density function of optimal tour lengths; 
\item
(7). Iteratively apply this approach, taking the expectation value of $(K-1)$-th iteration as the upper bound  of the $K$-th iteration, we have the expectation value after $K$ iterations ($K\geq 2$).

\end{itemize}


%
%
%
%
%


\textbf{LEMMA 1.} Applying Christofides algorithm to a ESTSP forms a truncated GB (TGB) for the probability density function of optimal tour lengths, with expectation (average) value at most 1.5OPT-$\epsilon$, where $\epsilon$ is a very small value.\\
\begin{proof}
This is because that Christofides algorithm assures that its result is at most 1.5OPT so that those tours with lengths more than 1.5OPT are excluded (truncated), as shown in Fig. 6 where tour lengths larger than 1.5OPT (1.5A) are truncated in black color.
\begin{figure} [htp!]

\begin{center}


{\includegraphics [width=0.4\textwidth,angle=-0] {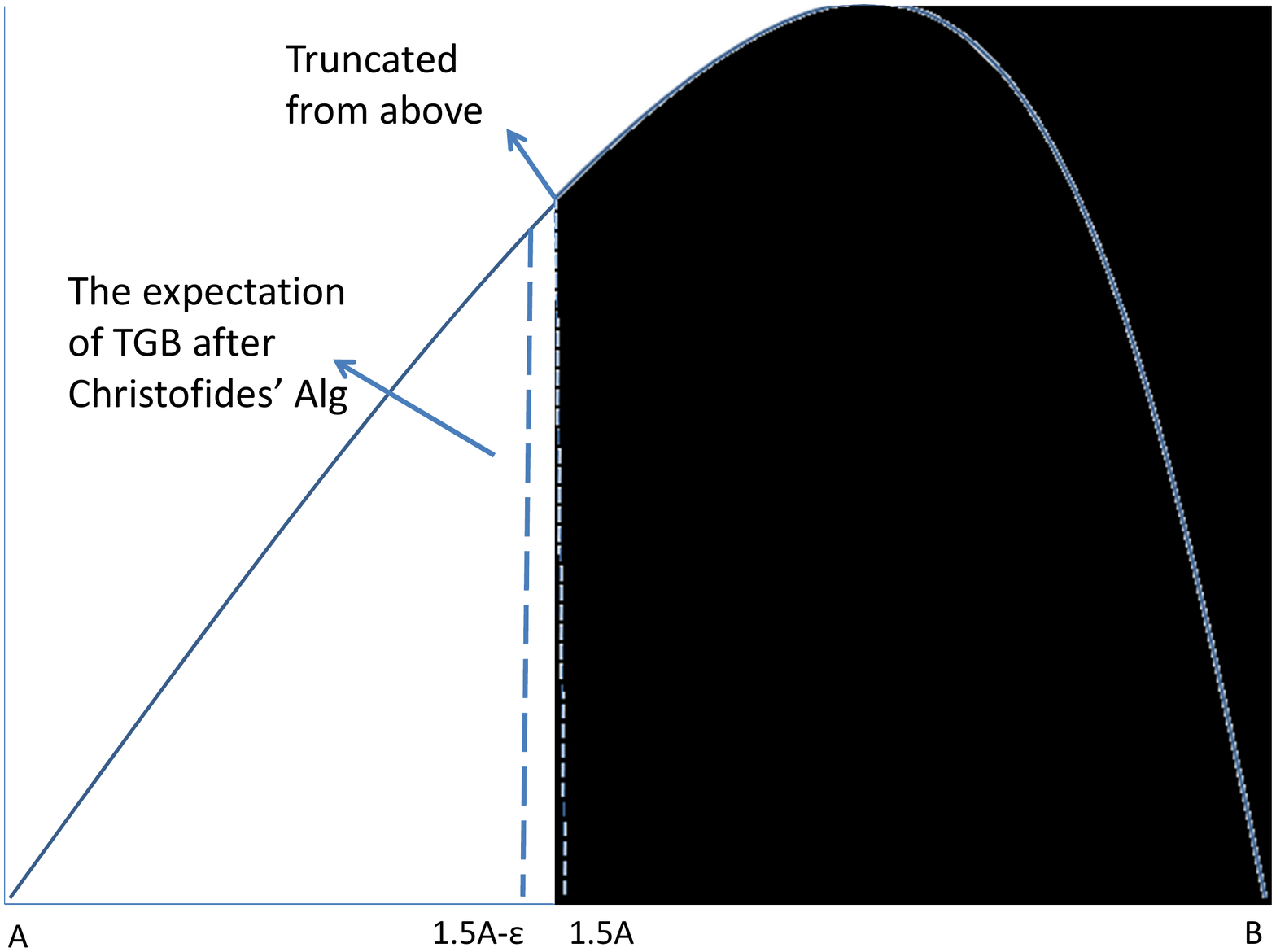}}


\caption{The Truncated GB by applying Christofides algorithm}

\end{center}

\end{figure}

The TGB in this case is truncated from above. Set $X$ as the variate of the GB, the probability density function (pdf) of TGB is given by
\begin{align}
f_t^1(x,\alpha,\beta, A, B,a,b)&=\frac{f(x,\alpha,\beta, A, B)}{Pr[a\leq X\leq b]} \nonumber\\
=&\frac{(x-A)^{\alpha-1}(B-x)^{\beta-1}} {\int_{a}^{b} (x-A)^{\alpha-1}(B-x)^{\beta-1}} \nonumber\\ \label{eq:TruncatedBetaDensity}
\end{align}
where $a$=$A$ and $b$=1.5$A$. Therefore 1.5$A$ is the upper bound.
We know that the average of Christofides algorithm is at most 1.5OPT-$\epsilon$ (set as $\mu_t^{1}$) where $\epsilon$ is a very small value, this is also validated in [Blaser et al., 2012].\\
\end{proof}
The four parameters of GB for some random TSP and TSPLIB instances are provided in Table 1 and 2 respectively. \\
\textbf{Definition 6.} $k$-opt method. Local search with $k$-exchange neighborhoods, also called $k$-opt, is the most widely used heuristic method for the TSP.
$k$-opt is a tour improvement algorithm, where in each step $k$ links of the current tour are replaced by $k$ links in such a way that a shorter tour is achieved (see [Helsgaun 2009] for detailed introduction).\\
In [Helsgaun 2009], a method with computational complexity of $O(k^3+k\sqrt{n})$ is introduced for $k$-opt.
\begin{figure} [htp!]

\begin{center}


{\includegraphics [width=0.4\textwidth,angle=-0] {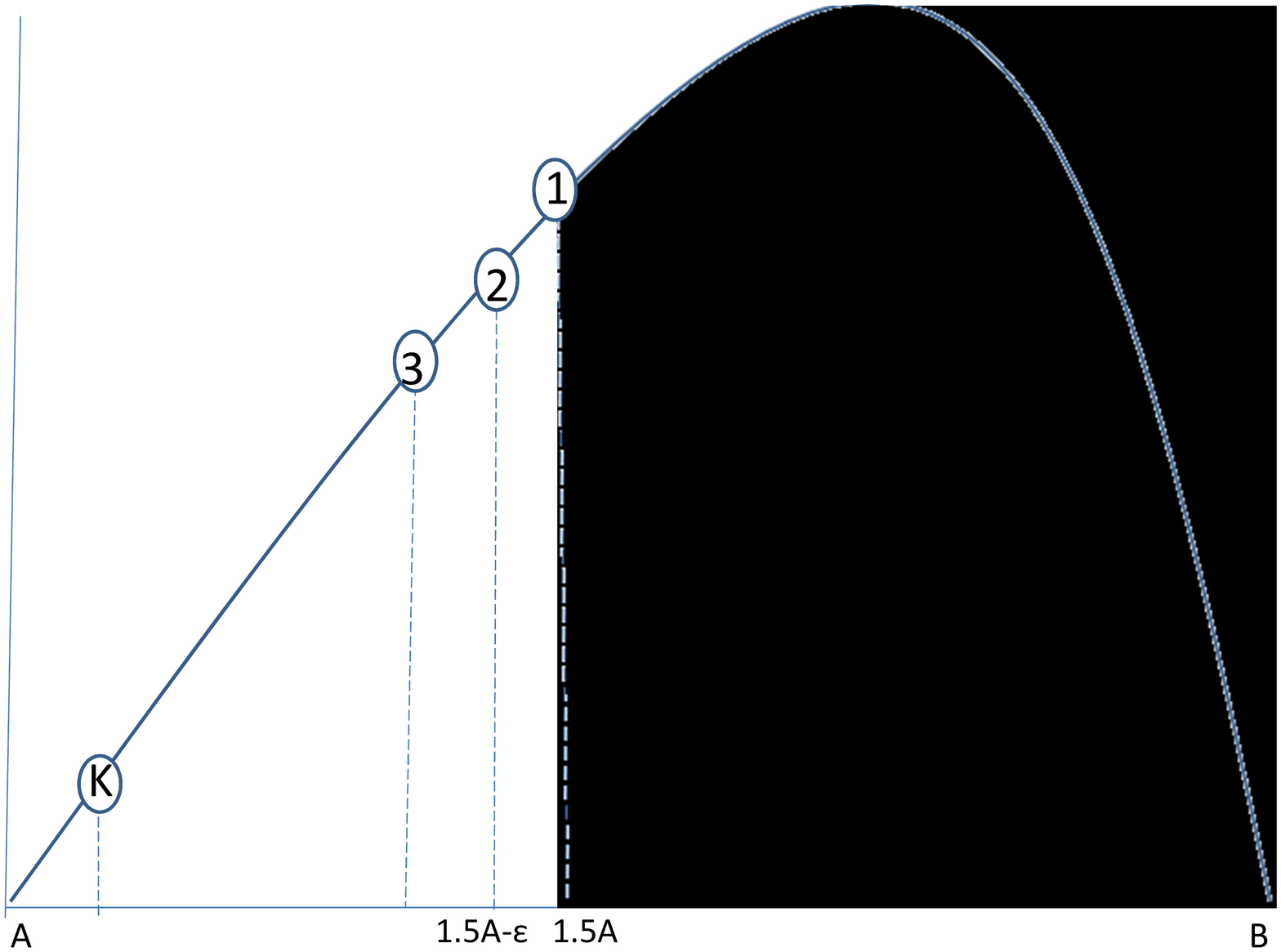}}


\caption{The Iteratively Truncated GB }

\end{center}

\end{figure}
\\
\textbf{LEMMA 2.} Applying $k$-opt to the result of Christofides algorithm forms another TGB for probability density function of optimal tour lengths.
\begin{proof}
Applying $k$-opt to the result obtained by Christofide algorithm as shown in Fig.7. The TGB in this case is truncated from above. Denote the first truncation by Christofides' algorithm as the first truncation ($K$=1). The probability density function of the second TGB is given by
\begin{equation}
f_t^2(x,\alpha,\beta, A, B,a_2,b_2)=\frac{(x-A)^{\alpha-1}(B-x)^{\beta-1}} {\int_{a_2}^{b_2} (x-A)^{\alpha-1}(B-x)^{\beta-1}}\label{eq:TruncatedBetaDensity}
\end{equation}
In this case, $a_2$=$A$, $b_2$=1.5$A$ because the distribution is based on the result after applying Christofides algorithm which assures the upper bound is at most 1.5$A$, see Fig.7.
Setting $\hat{x}$=$\frac{x-A}{B-A}$, $\hat{a_2}$=$\frac{A-A}{B-A}$=0, $\hat{b_2}$=$\frac{1.5A-A}{B-A}$=$\frac{0.5A}{B-A}$, we have
\begin{align}
C_0 &= {\int_{a_2}^{b_2} (x-A)^{\alpha-1}(B-x)^{\beta-1}}\mathrm{d}x \nonumber\\
&={\int_{0}^{\hat{b_2}} ((B-A)\hat{x})^{\alpha-1}((B-A)(1-\hat{x})^{\beta-1}}\mathrm{d}x \nonumber\\
&=(B-A)^{\alpha+\beta-1}B_2(0,\hat{b_2},\alpha,\beta)
\end{align}

where
\begin{equation}
B_2(0,t,\alpha,\beta)=\int_{0}^{t} x^{\alpha-1}(1-x)^{\beta-1} \mathrm{d}t \label{eq:TruncatedBy}
\end{equation}
By the definition of the expectation (mean) value (denoted as $\mu_t^2$) for $f_t^2(x,\alpha,\beta, A, B,a_2,b_2)$, we have
\begin{align}
\mu_t^{2}-A=\int_{a_2}^{b_2}(x-A)f_t^2(x,\alpha,\beta, A, B,a_2,b_2)\mathrm{d}x \nonumber\\
= \frac{ {\int_{a_2}^{b_2} (x-A)^{\alpha}(B-x)^{\beta-1}}\mathrm{d}x}{C_0} \nonumber\\
=  \frac{(B-A)^{\alpha+\beta}B_2(0,\hat{b_2},\alpha+1,\beta)}{C_0} \nonumber\\
= (B-A)\frac{B_2(0,\hat{b_2},\alpha+1,\beta)}{B_2(0,\hat{b_2},\alpha,\beta)} \nonumber\\
=>\mu_t^2=A+(B-A)\frac{B_2(0,\hat{b_2},\alpha+1,\beta)}{B_2(0,\hat{b_2},\alpha,\beta)}\label{eq:mu2}
\end{align}

Taking the expectation value of $(K-1)$-th iteration as the upper bound ($\hat{b_K}=\frac{\mu_t^{K-1}-A}{B-A}$) of the $K$-th iteration, we apply this approach Iteratively and have the expectation value after $K$ iterations ($K\geq 2$), denoted as $\mu_t^K$,
\begin{align}
\mu_t^K&=A+(B-A) \frac{B_2(0,\hat{b_K},\alpha+1,\beta)}{B_2(0,\hat{b_K},\alpha,\beta)} \nonumber \\
&= A+(B-A)g(\hat{b_K})\nonumber
\end{align}

\end{proof}










Next we provide the proof for our main theorem.\\
\textbf{THEOREM 1.} Applying TGB iteratively, we can obtain quality-proved approximation, i.e., (1+$\frac{1}{2}(\frac{\alpha+1}{\alpha+2})^{K-1}$)-approximation where $K$ is the number of iterations in TGB, $\alpha$ is the shape parameter of TGB and can be determined or estimated once TSP instance is given.
\begin{proof}
Notice that the expectation value of the ($K$-1)-iteration is taken as the upper bound ($\hat{b_K}=\frac{\mu_t^{K-1}-A}{B-A}$) of the $K$-iteration, as shown in Fig.7.
Setting

\begin{equation}
g(\hat{b_K})=\frac{B_2(0,\hat{b_K},\alpha+1,\beta)}{B_2(0,\hat{b_K},\alpha,\beta)}
\end{equation}
The exact expression of $g(\hat{b_K})$ can be stated in a hypergeometric series, and

\begin{equation}
B_2(0,\hat{b_K},\alpha,\beta)=\frac{\hat{b_K}^\alpha}{\alpha}F(\alpha,1-\beta,\alpha+1,\hat{b_K})
\end{equation}
and $ F(a,b,c,x)$

\begin{align}
&=1+\frac{ab}{c}x+\frac{a(a+1)b(b+1)}{c(c+1)2!}x^2 \nonumber \\
& +\frac{a(a+1)(a+2)b(b+1)(b+2)}{c(c+1)(c+2)3!}x^3+...
\end{align}
In all cases, we have $\alpha>$1, $\beta>$1, $\hat{b_K}\in (0,1)$, therefore $ F(a,b,c,x)$ is an monotonic decreasing function. We have
\begin{equation}
u_t^2= A+(B-A)g(\hat{b_2})\leq A+0.5A \frac{\alpha+1}{\alpha+2}
\end{equation}

continue this for $g(\hat{b_3})$, $u_t^3$, $g(\hat{b_4})$, $u_t^4$,..., so forth, we have
\begin{equation}
\hat{b_K}\leq \frac{0.5A}{B-A} (\frac{\alpha+1}{\alpha+2} )^{K-1}
\end{equation}
and 
\begin{align}
g(\hat{b_K}) &=\frac{B_2(0,\hat{b_K},\alpha+1,\beta)}{B_2(0,\hat{b_K},\alpha,\beta)} \nonumber \\
& \leq \frac{\alpha+1}{\alpha+2} \hat{b_K} \nonumber \\
& =\frac{0.5A(\frac{\alpha+1}{\alpha+2})^{K-1}}{B-A}, 
\end{align}
Therefore
\begin{align}
\mu_t^K&=A+(B-A) \frac{B_2(0,\hat{b_K},\alpha+1,\beta)}{B_2(0,\hat{b_K},\alpha,\beta)} \nonumber\\
&= A+(B-A)g(\hat{b_K})\nonumber\\
& \leq (1+\frac{1}{2}(\frac{\alpha+1}{\alpha+2})^{K-1})A, 
\end{align}

This means that, after the $K$-th iterative truncation, we can obtain the expectation value of $(\mu_t^{K})$ which is close to the optimum (OPT=A) when $K$ increases. Actually, to make the approximation less than $C_0$, the TGB algorithm needs $K$ to be at least (1+$\frac{log2(C_0-1)}{log(1-1/(\alpha+1))})$.
\end{proof}
Table 3 shows OPT,~$\alpha$,~$\beta$, iteration numbers and the approximation ratio (Appr) for TSPLIB instances with $n\leq 600$, where ($\alpha$,~$\beta$)  are obtained (or estimated) from ($A$, mean, variance, skewness) in Eqns (4)-(7), and $K$ is obtained in by TGB algorithm which modifies LKH code. We observe that the TGB results are consistent with LKH OPT results in most cases, there are only a few cases where TGB results are few percentage difference from OPT, with 0.2$\%$ off the true optimum on the average. For instance, the difference is $7.8\%$ for berlin52.tsp and $1.1\%$ for ulysses22.tsp. Our results are consistent with [Applegate 2003]. These results validate Theorem 1. Notice that LKH code performs very fast in practice and our results are based on average performance.
\begin{table}[ht]
\caption{Four parameters for some TSPLIB instances ($n \le$ 600)} 
\centering 
\begin{tabular}{c c c c c } 
\hline\hline 
&  $\alpha$  &K-1&(1+0.5($\frac{\alpha+1}{\alpha+2})^{K-1})A$ &Appr \\ [0.5ex] 
\hline 

ulysses22  & 17.52  & 91 & 82.17 & 1.0042\\
berlin52  & 57.40  & 101 & 9052.43 & 1.0900\\
pr76  & 115.54  & 2451 & 108159.41 & 1.0000\\
rat99  & 128.18  & 1821 & 1219.21 & 1.0000\\
kroA100  & 137.30  & 3366 & 21285.39 & 1.0000\\
pr299  & 422.28  & 29117 & 48194.85 & 1.0000\\
lin318  & 563.15  & 39112 & 42042.44 & 1.0000\\
rd400  & 735.15  & 34936 & 15275.79 & 1.0000\\
d493  & 695.93  & 129767 & 35018.32 & 1.0000\\
rat575  & 892.03  & 84814 & 6796.36 & 1.0000 \\[1ex]
\hline 
\end{tabular}
\label{table:nonlin} 
\end{table}

\section{Conclusions}

In this paper, for the first time, we proposed GB and Truncated Generalized Beta distribution (TGB) for the probability distribution of optimal tour lengths in a symmetric TSP in Euclidean space. 
Notice that our TGB results are based on expectation (average) value of probability distribution, which may be overestimated for the number of iterations. In practice, LKH algorithm performs very fast, with estimated computational complexity of $O(n^{2.2})$ [9].
A few possible research directions include:

\begin{itemize}

\item Improving the computational complexity. Currently the Christofides algorithm with minimum perfect matching has computational complexity $O(n^3)$. For large instances, this complexity should be reduced.

\item Find more efficient ways to compute especially the third and fourth central moments of a given TSP instance. 

\item Finding more applications.
With closed-form probability density function at hand, a lot of things can be done better. For instance, computing more statistical metrics, analyzing the average performance of approximation algorithms and others.
\end{itemize}





\textbf{Acknowledgement}

This research is partially supported by China National Science Foundation (CNSF) with project ID 61672136, 61650110513; and Xi Bu Zhi Guang Project (R51A150Z10). 
A version of manuscript is posted on http:\//arxiv.org\/pdf\/1502.00447.pdf


%













\end{document}